\documentclass[apj]{emulateapj}
\usepackage{graphicx}

\def\simgr{\,\hbox{\hbox{$ > $}\kern -0.8em \lower 1.0ex\hbox{$\sim$}}\,}
\def\simle{\,\hbox{\hbox{$ < $}\kern -0.8em \lower 1.0ex\hbox{$\sim$}}\,}

\slugcomment{}

\newcommand\xss{XSS J12270$-$4859}
\newcommand\fglone{3FGL J1544.6$-$1125}

\newcommand{\msp}{XMMU J083850.38$-$282756.8}
\newcommand{\fgl}{3FGL J0838.8$-$2829}
\newcommand{\fermi}{{\it Fermi}}

\def\chandra{{\it Chandra}}
\def\xmm{{\it XMM--Newton}}

\received{2017 June 14}
\accepted{2017 June 28}

\shortauthors{Halpern et al.}
\shorttitle{Redback Counterpart of 3FGL J0838.8$-$2829}

\begin{document}
\title{A Likely Redback Millisecond Pulsar Counterpart of 3FGL J0838.8$-$2829}

\author{J. P. Halpern\altaffilmark{1}, J. Strader\altaffilmark{2}, and M. Li\altaffilmark{1}}
\altaffiltext{1}{Columbia Astrophysics Laboratory, Columbia University,
550 West 120th Street, New York, NY 10027-6601, USA; jules@astro.columbia.edu}
\altaffiltext{2}{Center for Data Intensive and Time Domain Astronomy,
Department of Physics and Astronomy, Michigan State University,
East Lansing, MI 48824, USA}
\setcounter{footnote}{2}

\begin{abstract}
We obtained new optical observations of the X-ray source \msp,
the previously proposed counterpart of the $\gamma$-ray source \fgl. 
Time-series photometry in the $r^{\prime}$ band reveals periodic 
modulation of $\approx1$ magnitude that is characteristic of the heating
of the photosphere of a low-mass companion star by a compact object.
The measured orbital period is $5.14817\pm0.00012$~hr.  The shape of
the light curve is variable, evidently due to the effects of flaring and
asymmetric heating.  Spectroscopy reveals a companion of type M1 or later,
having a radial velocity amplitude of $315\pm17$~km~s$^{-1}$, with period
and phasing consistent with the heating interpretation.  The mass function
of the compact object is $0.69\pm0.11\,M_{\odot}$, which allows a neutron
star in a high-inclination orbit.  Variable, broad H$\alpha$ emission is
seen, which is probably associated with a
wind from the companion.  These properties, as well as the X-ray
and $\gamma$-ray luminosities at the inferred distance of $<1.7$~kpc,
are consistent with a redback millisecond pulsar in its non-accreting state.
A search for radio pulsations is needed to confirm this interpretation
and derive complete system parameters for modeling, although absorption by
the ionized wind could hinder such detection.
\end{abstract}

\keywords{gamma rays: stars --- pulsars: general
--- X-rays: individual (\msp)}

\section{Introduction}

A major achievement of the Large Area Telescope on the \fermi\
Gamma-ray Observatory is the detection of many recycled millisecond pulsars
(MSPs), some of which are accomplished with the help of prior radio ephemerides,
but most of them are new ones that are found in successful radio-pulsar
searches of its unidentified $\gamma$-ray source error circles.
Of the 205 pulsars detected by \fermi\ so far,
92 are recycled\footnote{https://confluence.slac.stanford.edu/display/GLAMCOG/
Public+List+of+LAT-Detected+Gamma-Ray+Pulsars}.
Most interesting among the \fermi\ MSP discoveries are the black widow
(BW) and ``redback'' binary systems \citep{rob13}.
BWs have companions of substellar mass, while redbacks
have $>0.1\,M_{\odot}$, bloated companions that are usually close
to filling their Roche lobes.  Both BWs and redbacks are compact binaries,
with orbital periods $\simle$1~day.  Their properties
connect BWs and redbacks to their long-supposed low-mass X-ray binary (LMXB)
progenitors, which spin them up by accretion \citep{alp82}.  A direct link
was forged by the three redbacks that have been observed to transition
between radio pulsar and accreting states on timescales of years:
PSR J1023+0038 \citep{arc09}, \xss\ \citep{bas14,roy15},
and PSR J1824$-$2452I in the globular cluster M28 \citep{pap13}.  
Currently, $\approx60$ BW and redback pulsars are known that
are almost equally divided between globular clusters and
Galactic field populations\footnote{https://apatruno.wordpress.com/about/millisecond-pulsar-catalogue/}.

In addition to the BW and redback systems that are certified radio MSPs,
it has become possible through X-ray and optical follow-up of \fermi\ source
error circles to identify counterparts that are almost certainly MSP binaries, even without
obtaining a radio-pulsar detection.  Their distinctive optical and X-ray light curves,
spectroscopic orbital parameters, and positional coincidence with a $\gamma$-ray source
allow some objects to be classified as BW or redback pulsars whose spin
parameters are not yet known.  \fermi\ sources that have been identified
in this way are 3FGL J0212.1+5320 \citep{li16,lin17},
1FGL J0523.5$-$2529 \citep{str14},
2FGL J1311.7$-$3429 \citep{kat12,rom12a,rom12b,rom15b},
2FGL J1653.6$-$0159 \citep{kon14,rom14},
3FGL J2039.6$-$5618 \citep{rom15a,sal15}, and
1FGL J2339.7$-$0531 \citep{rom11,kon12}.
Radio pulsations at 2.88~ms were subsequently detected from
1FGL J2339.7$-$0531/PSR J2339$-$0533
by \citet{ray14}, and 2.56~ms pulsations were found in $\gamma$-rays
from 2FGL J1311.7$-$342/PSR J1311$-$3430 \citep{ple12} with
the help of its optical orbital ephemeris.

Complementing the above list of sources, 3FGL J0427.9$-$6704 \citep{str16},
and \fglone\ \citep{bog15b} have X-ray/optical counterparts that
appear to be transitional MSPs in the accreting state.
A related discovery is 3FGL J1417.5$-$4402/PSR J1417$-$4402,
with a giant companion in a 5.4~day orbit \citep{str15}
and 2.66~ms radio pulsations \citep{cam16}.
It is thought to be a typical MSP observed in the late stages of recycling,
one that will end its evolution with a white dwarf companion in a several
day orbit. The difficulty of detecting radio pulsations over most
of the orbit of PSR J1417$-$4402, as well as from most redbacks, 
is apparently due to the absorption by the winds from their companions.
This suggests that some $\gamma$-ray pulsars may be completely enshrouded
most of the time \citep[as proposed by][]{tav91}
and will remain as only putative MSPs unless their pulsations
can be found in X-rays or $\gamma$-rays.

Several authors have performed systematic analyses of the spectral
and temporal properties of unidentified \fermi\ sources using machine
learning techniques to assign the tentative classifications as
active galactic nuclei, young pulsars, or MSPs
\citep{ack12,lee12,mir12,mir16,saz16}.  In particular,
\citet{mir16} classified \fgl\ among the high-confidence MSPs.
In \citet[][Paper I]{hal17}, we identified a candidate binary MSP counterpart
for \fgl\ from an \xmm\ X-ray and optical study of its error circle
(see also \citealt{rea17}),
as well as in out own ground-based optical data.
In this paper, we report on new optical and radio observations that
characterize the source, \msp.  Section~2 describes time-series photometry
that determines its 5.15~hr orbital period and characterizes the orbital
light curve.  Section~3 reports spectroscopy that yields a radial velocity
curve for the secondary, and reveals variable, broad H$\alpha$ emission lines.
Section~4 presents tentative inferences about the binary parameters, distance,
and luminosity, and discusses the heating light curve and flaring activity
and possible interpretations of the  H$\alpha$ emission line.
Section~5 summarizes the main conclusions.

\section{Time-Series Photometry}

\begin{deluxetable}{llc}
\tablewidth{0pt}
\tablecaption{Log of Time-series Photometry\tablenotemark{a}}
\tablehead{
\colhead{Telescope/Instrument} & \colhead{Date (UTC)} & \colhead{Time (UTC)}
}
\startdata
MDM 2.4 m/OSMOS   & 2016 Dec 24     & 07:27--12:29 \\
MDM 2.4 m/OSMOS   & 2016 Dec 27     & 07:02--12:41 \\
LCO 1 m/Sinistro  & 2017 Jan 2      & 02:00--07:56 \\
MDM 2.4 m/OSMOS   & 2017 Jan 28     & 05:01--09:57 \\
MDM 2.4 m/OSMOS   & 2017 Feb 25     & 02:57--08:20 \\
LCO 1 m/Sinistro  & 2017 Mar 27--28 & 23:27--05:33   
\enddata
\tablenotetext{a}{All exposures are 300~s in the $r^{\prime}$ filter.}
\label{tab:optlog}
\end{deluxetable}

\begin{figure}
\centerline{
\includegraphics[angle=0,width=1.\linewidth,clip=]{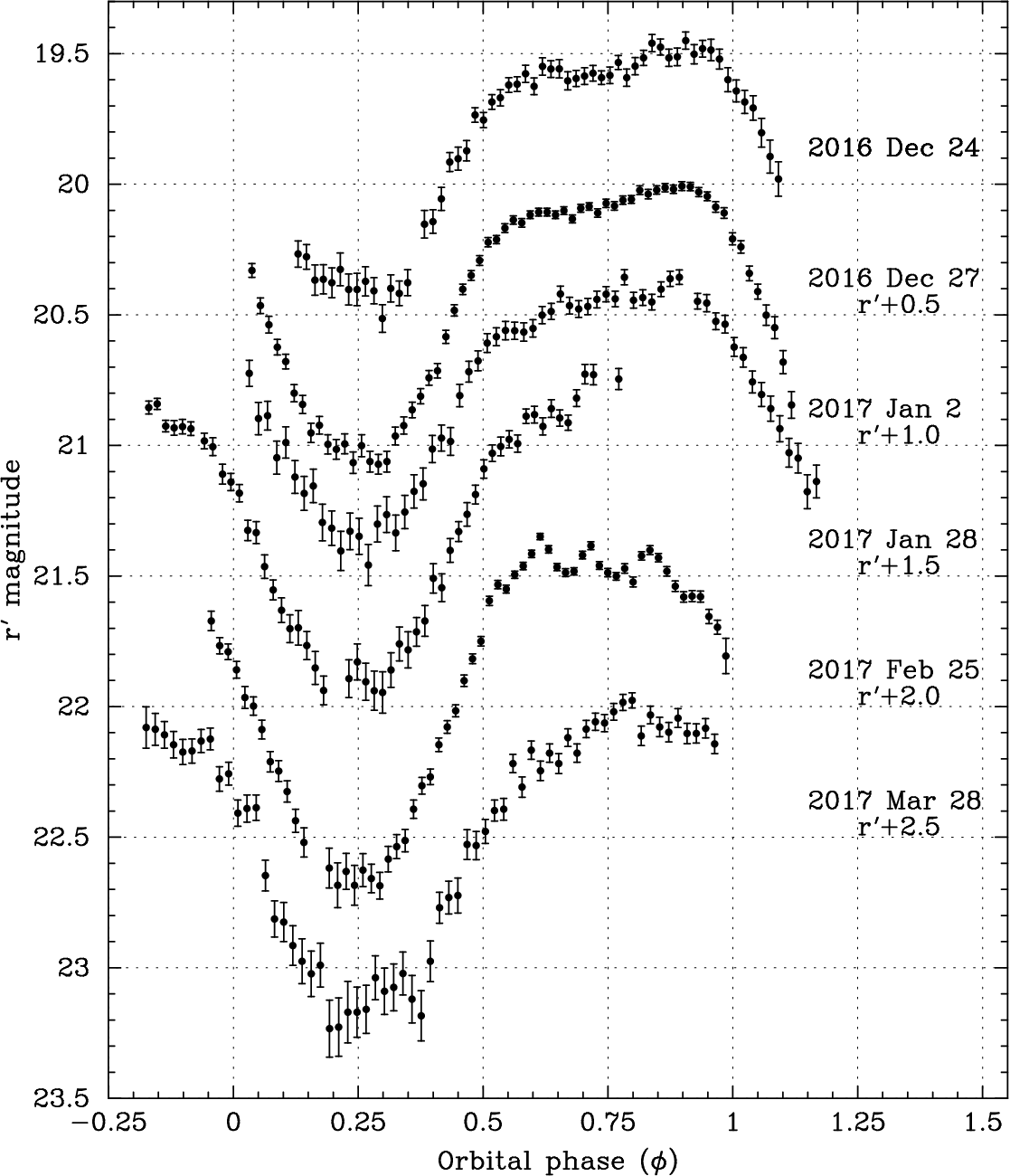}
}
\caption{Optical light curves as a function of orbital phase according
to the ephemeris in Table~\ref{tab:ephem}.  A log of the observations is
given in Table~\ref{tab:optlog}.  Each data set after 2016 December 24
is displaced downward by a multiple of 0.5 mag for clarity.
}
\label{fig:optical_phase}
\end{figure}

Over 3 months in 2016--2017 we obtained six nights of time-series photometry
of \msp, the MSP candidate for \fgl\ from Paper~I and \citet{rea17}
using either the MDM Observatory
2.4m Hiltner telescope and Ohio State Multi-Object Spectrograph \citep[OSMOS;][]{mar11} 
in imaging mode, or a robotic 1m telescope at the CTIO node of the Las Cumbres Observatory (LCO)
with its Sinistro imager \citep{bro13}.  A log of the observations is given in
Table~\ref{tab:optlog}.  An SDSS $r^{\prime}$ filter was used in
each case with an exposure time of 300~s.  Run lengths ranged between
5 and 6 hours.  The goals of this single-filter observing program were
to obtain a precise orbital ephemeris, to monitor for rapid flaring
as was seen in one of the \xmm\ observations (Paper~I), and to characterize
any other variability of the light curve.

The optical position listed in Table~\ref{tab:ephem} was derived from a 2.4m
image using stars from the USNO B1.0 catalog \citep{mon03} for an astrometric solution,
and is consistent with the optical position given in Paper~I.  The $0.\!^{\prime\prime}3$
error indicated in each coordinate includes the nominal uncertainty of the catalog
coordinates.

Differential photometry and magnitude calibration were performed with
respect to a calibrated comparison star.  Images were
inspected for cosmic-ray contamination, and a few points were rejected
for cosmic-ray hits.  The six resulting light curves are shown in
Figure~\ref{fig:optical_phase} as a function of orbital phase,
the determination of which is described below.  
Here we adopt the convention in which phase zero
is the ascending node of the pulsar, $\phi=0.25$ is the inferior
conjunction of the companion, and $\phi=0.75$ is the superior conjunction.
The shape of the light curve is characteristic of the heating of the side
of the companion facing the pulsar, with a maximum brightness
of $r^{\prime}\approx19.5$ and a minimum of $r^{\prime}\approx20.6$.  

There is occasional flaring, which is most easily visible in the 2017 January~28
and February~25 light curves in Figure~\ref{fig:optical_phase}.  A quiescent
state, if one can be recognized in the data, is perhaps best represented
by the smooth light curve on 2016 December~27.  It shows a sloping
maximum that is not symmetric about the expected phase 0.75 of peak heating,
but is rising up to $\phi=0.90$.  Such behavior is not uncommon among redbacks
\citep{wou04,li14,den16}, indicating that the heating of the companion is
frequently not symmetric about the line connecting the stars.

In order to derive an orbital ephemeris in the presence of flaring and other
possible sources of variable asymmetry of the light curve, we use
the epoch of minimum in the light curve as $\phi=0.25$, the fiducial phase,
presuming that the timing of the minimum would be least affected by
variability.  Mid-times of each exposure were converted to barycentric dynamical
time (TDB) using the utility of \citet{eas10}.  Approximately 25 points around
each minimum were fitted to a quadratic, and the calculated epochs of minimum
were fitted to a constant orbital period, $P_{\rm orb}=5.14817\pm0.000012$~hr.
The precise parameter values are listed in Table~\ref{tab:ephem}, with the epoch
of pulsar ascending node $T_0$ taken to be at $\phi=0$.  The average residual
of the epochs of minimum from the fitted ephemeris is $\approx 2$~minutes.
For the ephemeris presented here we did not use the first observation
(2016 December 24) as its minimum was poorly determined due to bad seeing,
although its inclusion does not change the results within their errors.

\begin{deluxetable}{lc}
\tablewidth{0pt}
\tablecaption{Photometric Orbital Ephemeris}
\tablehead{
\colhead{Parameter} & \colhead{Value}
}
\startdata
R.A. (J2000)              & $08^{\rm h}38^{\rm m}50^{\rm s}\!.44(2)$ \\
Decl. (J2000)             & $-28^{\circ}27^{\prime}57^{\prime\prime}\!.3(3)$ \\
Time span (MJD)           & 57746--57840 \\
Epoch $T_0$ (MJD TDB)\tablenotemark{a}      & 57781.2524(8) \\
Orbital period $P_{\rm orb}$ (day)  & 0.214507(5) 
\enddata
\label{tab:ephem}
\tablenotetext{a}{Epoch of the ascending node of the putative pulsar
$\phi=0$ in Figure~\ref{fig:optical_phase}.}
\end{deluxetable}

\section{Optical Spectroscopy}

We performed spectroscopy on four different nights, from 2017 February 3 to April 24 (UT), with the Goodman spectrograph \citep{cle04} on the Southern Astrophysical Research (SOAR) telescope. On the first night spectra were taken with a 1200~l~mm$^{-1}$ grating and a $1.\!\arcsec03$ wide slit over the wavelength range of 7700--8700~\AA, yielding a resolution of 1.6 \AA. The remaining observations all used a 400~l~mm$^{-1}$ grating and a $1.\!\arcsec03$ slit over the wavelength range 4800--8830~\AA, giving a resolution of 5.4~\AA. The exposure times ranged from 900 to 1500~s, depending on the brightness of the source and the seeing. The spectra were reduced in the standard manner.  Molecular bands indicate that the companion star is a late K or early M type, probably M0--M1 on the night side, and slightly hotter on the illuminated side.

We derived barycentric radial velocities through cross-correlation with bright stars taken with the same setup. Generally we used the region of the \ion{Ca}{2} near-infrared triplet, but we substituted the region around Mg$b$ for spectra with low signal-to-noise in the \ion{Ca}{2} triplet region. The respective epochs for each spectrum are given as Modified Julian Day (MJD) in the TDB system. Given the faintness of the star, the formal radial velocity uncertainties are themselves uncertain, but the final results are nevertheless robust (see below). The radial velocity data are listed in Table~\ref{tab:velocity}.

\begin{deluxetable}{lcccc}
\tablewidth{0pt}
\tablecaption{Barycentric Radial Velocities}
\tablehead{UT Date & MJD (TDB) & Exposure & $v_r$ & $\sigma(v_r)$ \\
&  & (s) & (km s$^{-1}$) & (km s$^{-1}$)}
\startdata
2017 Feb 3  & 57787.1650 & 900   & 342.3   & 25.6\\
2017 Feb 3  & 57787.1756 & 900   & 316.8   & 25.0\\
2017 Feb 12 & 57796.1260 & 1200  & 339.2   & 25.5\\
2017 Feb 12 & 57796.1444 & 1200  & 465.9   & 22.5\\
2017 Mar 28 & 57840.1192 & 1200  & 436.0   & 28.3\\
2017 Mar 28 & 57840.1332 & 1200  & 431.3   & 37.7\\
2017 Mar 28 & 57840.1513 & 1200  & 407.9   & 27.1\\
2017 Mar 28 & 57840.1653 & 1200  & 377.1   & 32.4\\
2017 Mar 28 & 57840.1839 & 1200  & 125.4   & 25.7\\
2017 Apr 24 & 57867.0436 & 1200  & --222.4 & 21.2\\
2017 Apr 24 & 57867.0576 & 1200  & --184.7 & 23.5\\
2017 Apr 24 & 57867.0756 & 1200  & --117.3 & 24.0\\
2017 Apr 24 & 57867.0896 & 1200  & 20.2    & 21.6\\
2017 Apr 24 & 57867.1269 & 1500  & 289.0   & 20.5
\enddata 
\label{tab:velocity}
\end{deluxetable}

\begin{figure}
\centerline{
\includegraphics[angle=270,width=0.97\linewidth,clip=]{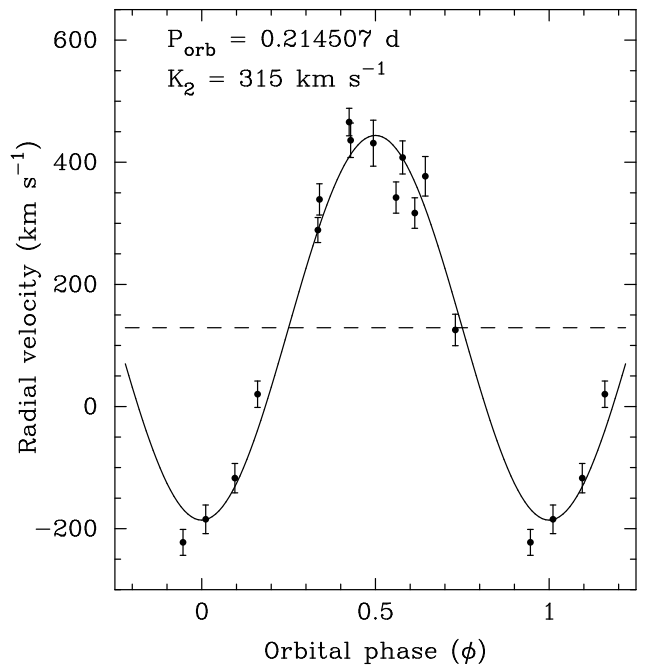}
}
\caption{Spectroscopic radial velocities from Table~\ref{tab:velocity}
and best-fit sinusoid assuming the photometric orbital ephemeris of
Table~\ref{tab:ephem}.  The systemic velocity (dashed line) is
$v_{\rm sys}=129\pm15$ km~s$^{-1}$.
}
\label{fig:velocity}
\end{figure}

Since the orbital period and time of ascending node are determined with more precision from the photometry than is possible from the presently limited spectroscopy, we adopted the photometric values of $P_{\rm orb}$ and $T_0$ from Table~\ref{tab:ephem}, nonetheless finding that the photometric values fall within the uncertainties resulting from an independent fit to the radial velocities only.  Given the short period and presumed long history of the system, we also assume zero eccentricity. Hence, the only free parameters are the velocity semi-amplitude of the secondary $K_2$ and the systemic velocity $v_{\rm sys}$. Using a circular Keplerian fit, we find best-fit values of $K_2 = 315\pm17$ km~s$^{-1}$ and $v_{\rm sys} = 129\pm15$ km~s$^{-1}$, with the uncertainties determined via bootstrap. This immediately yields a mass function of $0.69\pm0.11 M_{\odot}$ for the compact object.  If we instead assume that all radial velocities have the same uncertainties, the best-fit $K_2$ and $v_{\rm sys}$ do not change. The best-fit radial velocity curve is shown in Figure~\ref{fig:velocity}.

\begin{figure}
\centerline{
\includegraphics[angle=270,width=0.95\linewidth,clip=]{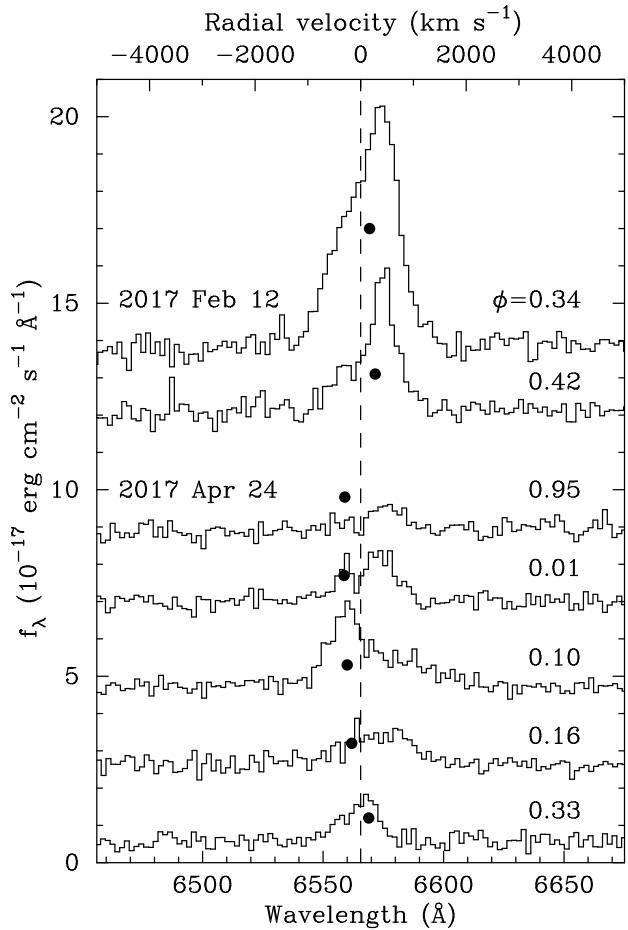}
}
\caption{Selected spectra, two from 2017 February 12 and five from
2017 April 24, showing a broad H$\alpha$ emission line.  
Spectra have been shifted upward in steps of
$2\times10^{-17}$ erg~cm$^{-2}$~s$^{-1}$~\AA$^{-1}$ for clarity.
The vertical dashed line represents the wavelength of H$\alpha$ at
the systemic velocity of the binary.
The binary orbital phase according to the ephemeris of
Table~\ref{tab:ephem} is labeled. The filled circles indicate
for each spectrum the fitted radial velocity of the companion
relative to the center-of-mass (systemic) velocity.
}
\label{fig:specplot}
\end{figure}

In addition to the radial velocities, a notable feature of the optical spectra is the variable Balmer-line emission. The spectra on 2017 February~12 show remarkably broad, asymmetric H$\alpha$ emission (this is the only night where H$\beta$ emission, which is similarly broad, is clearly present). On 2017 March~28 the H$\alpha$ emission is present, but weak. In the last set of data, from 2017 April~24, the emission is stronger, though still not as strong as on 2017 February~12.  Figure~\ref{fig:specplot} shows the H$\alpha$ emission lines on the days when it was strongest.  The maximum observed equivalent width was $\approx90$~\AA.  The FWHM ranges from 550--1510 km~s$^{-1}$, accounting for the spectral resolution. The line is generally asymmetric and sometimes double-peaked, with peak separations of 520--770 km~s$^{-1}$.  It is difficult to trace the velocities of the emission line due to the changing strengths of its multiple components.  While they are shifted from the absorption-line radial velocities, there is no clear pattern with respect to orbital phase.  However, there is often a peak in the line at or near the radial velocity of the companion, as marked in Figure~\ref{fig:specplot}.

\section{Discussion}

\subsection{Binary Parameters, Distance, and Luminosity}

Assuming that the intrinsic spectral type of the companion star
is M1 or later (because some of the heated side may still
be visible at minimum light), its mass and radius are
$<0.53\,M_{\odot}$ and $<0.49\,R_{\odot}$, respectively \citep{pec13}.
For these values, and assuming a $1.6\,M_{\odot}$ neutron star
(to allow for some accretion),
the orbital separation is $<1.94\,R_{\odot}$ and the Roche-lobe radius
of the companion is $<0.56\,R_{\odot}$ \citep{egg83},
consistent with the star almost filling its Roche lobe.
Alternatively, the companion may be slightly bloated compared
to a main-sequence star of the same mass,
as has been found in several studies
of redbacks \citep{li14,bel16,lin17}.

The inclination of the orbit must be substantial in order
to account for the observed optical modulation.  The mass function
of $0.69\pm0.11 M_{\odot}$ places some limits on the inclination
for reasonable values of the stellar masses.
Assuming a companion mass in the range of $0.3<M_c<0.5\,M_{\odot}$
and a neutron star of $1.4<M_{\rm ns}<2\,M_{\odot}$,
the inclination angle must be $46^{\circ}<i\le90^{\circ}$.

At minimum light, $r^{\prime}\approx20.6$ corresponds to a
distance $<1.7$~kpc for a star of type M1 or later,
including the Galactic extinction in this direction of $A_{r^{\prime}}=0.38$
\citep{sch11}.  The 0.3--10~keV X-ray luminosity of the source is
$2\times10^{31}\,(d/1\,{\rm kpc})^2$ erg~s$^{-1}$ (Paper~I),
which is in the range of BWs or redbacks in the radio-pulsar
state \citep{rob15}. 
In contrast, accreting redbacks have an average
$L_x\sim3\times10^{33}$ erg~s$^{-1}$ \citep{bog15a},
with even brighter flares.  The 0.1--100~GeV
luminosity of \fgl, $1.5\times10^{33}\,(d/1\,{\rm kpc})^2$
erg~s$^{-1}$ \citep{ace15}, is compatible with the pulsed emission
having $\sim10\%$ of the spin-down power of the putative pulsar,
an average efficiency observed among $\gamma$-ray MSPs \citep{abd13}.

\subsection{Heating Light Curve}

Because its spectral type, hydrogen emission, and moderate ($\approx1$~mag)
orbital variation suggest that it has a nearly Roche-lobe filling main-sequence
secondary, the optical properties of \msp\ are consistent with
those of a redback MSP in its ``radio-pulsar'' (non-accreting) state.
Its light curve resembles that of the brighter 4.75~hr
redback PSR J1023+0038 \citep{wou04,tho05} and even more so the 6.01~hr
redback PSR J1048+2339 \citep{den16}, the latter being quite similar
to \msp\ in its apparent magnitude, $\approx1$~mag orbital modulation, and
occasional bright flares.
In contrast, a BW, with its substellar companion, sometimes
deficient in hydrogen and usually underfilling its Roche lobe,
is modulated by several magnitudes around the orbit, e.g.,
the 94~minute binary PSR J1311$-$3430 \citep{rom12b,rom15b}
and others \citep{bre13,tan14,rom16a}.

The asymmetry of the heating light curve of \msp\ recalls
the similar behavior of PSR J1023+0038 \citep{wou04,bog11} in
optical and X-rays, as well as PSR J1048+2339 in optical
\citep{den16}.  Both of these pulsars have sloping maxima,
which are highest at $\phi\approx0.6$ and decrease toward $\phi\approx1.0$.
The only difference in \msp\ is that the slope is reversed,
at least in its apparently quiescent state,
increasing from $\phi\approx0.5$ to $\phi\approx0.9$
instead of decreasing. \citet{rom16b} have attempted to
model such asymmetries using the effect of the orbital motion
on the shape of an intrabinary shock, but the impact on
the optical heating light curve is much smaller than what
we observe, requiring an additional physical mechanism
to explain such extreme behavior.

\citet{rom16b}, as well
as \citet{tan14} and \citet{li14}, suggest that direct
channeling of the pulsar wind by magnetic fields intrinsic
to the companion star may produce asymmetric heating patterns.
\citet{san17} have begun to model such heating, and
it appears that their offset dipole magnetic field geometry
produces a light curve that resembles the data.
Such strong magnetic fields can be expected in the tidally locked,
rapidly rotating stars in BW and redback binaries.  This also
implies that large starspots could be present whose intrinsic
brightness distribution could be responsible for light-curve
asymmetries and variations. \citet{van16} found evidence for
such spots during intensive monitoring of the redback
PSR J1723$-$2737, which sometimes shows a periodic component
slightly displaced from the orbital period, implying that the
companion's rotation is not exactly tidally locked.

\subsection{Optical and X-ray Flares}

\begin{figure}
\centerline{
\includegraphics[angle=270,width=0.98\linewidth,clip=]{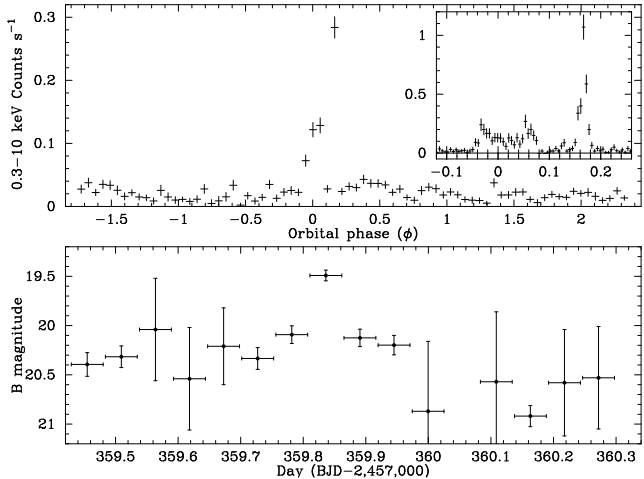}
}
\caption{Background-subtracted \xmm\ light curves, adapted from Paper I.
Top: EPIC pn 0.3--10~keV light curve in 1000~s bins
on 2015 December 2 (ObsID 0790180101) as a function
of orbital phase extrapolated from the ephemeris in Table~\ref{tab:ephem}.
The integer phase corresponds to the ascending node of the putative pulsar.
The inset shows the 1.2~hr flaring episode at a higher resolution (100~s bins).
Bottom: simultaneous \xmm\ OM magnitudes from 4400~s exposures in $B$-band.
}
\label{fig:xray_phase}
\end{figure}

Definite flaring behavior is present during some of the observations,
most notably during $0.6<\phi<0.85$ on 2017 January 28 and February 25.
In the BW PSR J1311$-$3430, bright flares have been seen in optical
and X-rays, which, because they appear to exceed the spin-down power
of the pulsar, could be coming from stored energy
in the companion's magnetic field \citep{rom15b}.
The same may be true of the occasional flares seen from
PSR J1048+2339 in the Catalina Real Time Transient Survey
\citep{den16}, which are difficult to account for assuming
isotropic pulsar-wind heating.  We do not know the pulsar parameters
for \msp, so it is not yet possible to make such a
quantitative comparison for it.

\msp\ is the first non-accreting MSP binary in which a simultaneous
X-ray and optical flare was observed (Paper~I).
The optical ephemeris of Table~\ref{tab:ephem} is precise enough to extrapolate
to the epoch of the \xmm\ light curve of 2015 December 2 (ObsID 079018010),
enabling us to examine the orbital dependence of the X-ray variability. 
The phase uncertainty of the extrapolation is $\sigma(\phi)\approx0.05$.
Figure~\ref{fig:xray_phase} is a reproduction of Figure~10 from Paper~I,
now plotted as a function of orbital phase.  The dramatic 1.2~hr long flare
spans $-0.05<\phi<0.19$, with peaks at $\phi=-0.03,0.06$, and 0.17.
The flare's final and highest peak has a maximum luminosity of
$\approx1\times10^{33}\,(d/1\,{\rm kpc})^2$ erg~s$^{-1}$
in a single 100~s bin.  Since it occurs close to $\phi=0.25$,
either the flaring region must not be very close to the
heated photosphere where it could be occulted,
or the inclination angle of the orbit must not be very large.
In redbacks there is often a broad dip in the X-ray emission centered
around $\phi=0.25$, which is attributed to the occultation
of an intrabinary shock that is very close to the companion
\citep{bog14a,bog14b,gen14,hui15,rob15,rom16b}.  There is no evidence for such
orbital modulation of the ``quiescent'' emission in
Figure~\ref{fig:xray_phase},
which instead may be dominated by low-level flaring.

The simultaneous flare in the \xmm\ optical monitor
(Figure~\ref{fig:xray_phase}, bottom), for which $B=19.5$ averaged over the
4400~s exposure time is an increase of $\approx1.5$~mag over the
quiescent level, is more intense than any of the smaller
flares that we have seen in 33 hours of $r^{\prime}$-band photometry.
This implies either that the flares are very blue, or that emission at the level
of the \xmm\ flare is relatively rare.

\subsection{H$\alpha$ Emission}

The broad width of the H$\alpha$ emission line suggests that it is not associated with the chromosphere of the companion. Instead, it could be due to a wind driven from the heated side of the companion by the impact of the pulsar wind, or by the indirect radiation from an intrabinary shock. Alternatively, it could be associated with an accretion disk around the primary, although we disfavor this scenario because there is no other signature of a disk, e.g., a hot optical continuum.  Limited by the small number of spectra collected, we see no definite pattern in velocity that could locate the origin of the line emission.  However, it appears that often, but not always, there is a peak in the line profile at or near the radial velocity of the companion, which suggests that a wind is being launched from the star. \citet{rom15b} found similar, variable \ion{He}{1} emission lines in PSR J1311$-$3430, which they attributed to a wind from the companion.

It may be relevant that the strongest H$\alpha$ emission was observed
on 2017 February 12, between the January 28 and February 25 light curves
when the optical continuum flaring was strongest.  In contrast, on March 28
the H$\alpha$ was weakest, while the simultaneous continuum light curve
fortuitously obtained on that night showed no flaring.
In any case, the large changes in equivalent width and profile suggest
that a variable stellar wind is involved in producing the H$\alpha$ emission.

\section{Conclusions and Further Work}

A photometric and spectroscopic study of the previously suggested
X-ray counterpart of the \fermi\ source \fgl\ reveals a 5.15~hr
orbit with a heating light curve, and an M dwarf companion
characteristic of a redback MSP system.  X-ray and $\gamma$-ray
fluxes are compatible with an MSP identification at a typical distance
and spin-down power for \fermi\ MSPs.  Following on similar
discoveries of this distinctive class of MSP binary in positional
coincidence with $\gamma$-ray sources, we conclude that \msp\
is the redback counterpart of \fgl\ even though its pulsations
have not yet been detected.  This source adds to a growing
sample of redbacks that have asymmetric heating light curves
and strong flaring activity in X-ray and optical that are
not well fitted by existing models of heating meditated by
an intrabinary shock between the pulsar wind and the companion's
wind.  This suggests that magnetic fields intrinsic to the companion
shape the light curve, both by channeling the pulsar wind directly
to its photosphere and by tapping its own energy in transient
reconnection events.

Variable H$\alpha$ emission may be coming from the wind driven
from the companion, which could be responsible for the absence
(so far) of detected radio pulsations due to free-free absorption
or dispersion.  Nevertheless, sensitive searches are warranted for
radio pulsations at a number of epochs and at all orbital phases,
since the uncertain configuration and possibly variable column
density of the absorbing plasma may allow rare windows of
transparency.  A search of the \fermi\ $\gamma$-rays
for pulsations may also be feasible with the aid of the optical
ephemeris, to be refined in the future.

\section{Acknowledgements}

We thank the anonymous referee for excellent suggestions.
Jessica Klusmeyer obtained two of the optical
light curves at the MDM Observatory.
MDM Observatory is operated by Dartmouth College,
Columbia University, the Ohio State University, Ohio University,
and the University of Michigan.
This work makes use of observations from the LCO network and the
SOAR telescope, which is a joint project of the Minist\'{e}rio da Ci\^{e}ncia,
Tecnologia, e Inova\c{c}\~{a}o (MCTI) da Rep\'{u}blica Federativa do Brasil,
the U.S. National Optical Astronomy Observatory (NOAO),
the University of North Carolina at Chapel Hill (UNC),
and Michigan State University (MSU).
Support for this work was provided by the National Aeronautics and
Space Administration through \chandra\ Award Number SAO GO6-17027X
issued by the \chandra\ X-ray Observatory Center,
which is operated by the Smithsonian Astrophysical Observatory
for and on behalf of the National Aeronautics Space Administration
under contract NAS8-03060.
J.S. acknowledges support from NASA grant NNX15AU83G and a Packard Fellowship.
This investigation also uses observations obtained with \xmm,
an ESA science mission with instruments and contributions directly
funded by ESA Member States and NASA.

\end{document}